\documentclass[12pt,a4paper]{article}
\begin{document}

\begin{flushright}
DPNU-00-18
\end{flushright}
\vspace{30pt}

\begin{center}
{\LARGE \scshape Berry Connections and }

\medskip {\LARGE \scshape Induced Gauge Fields in }

\medskip {\LARGE \scshape Quantum Mechanics on Sphere \vspace{30pt}}

Hitoshi IKEMORI\footnote{%
ikemori@biwako.shiga-u.ac.jp} \\[0pt]
\textit{Faculty of Economics, Shiga University,}

\textit{Hikone, Shiga 522-8522, Japan}

\medskip

Shinsaku KITAKADO\footnote{%
kitakado@eken.phys.nagoya-u.ac.jp} \\[0pt]
\textit{Department of Physics, Nagoya University,}

\textit{Nagoya 464-8602, Japan}

\medskip

Hideharu OTSU\footnote{%
otsu@vega.aichi-u.ac.jp} \\[0pt]
\textit{Faculty of Economics, Aichi University, }

\textit{Toyohashi, Aichi 441-8522, Japan}

\medskip

Toshiro SATO\footnote{%
tsato@matsusaka-u.ac.jp} \\[0pt]
\textit{Faculty of Policy Science, Matsusaka University,}\\[0pt]
\textit{\ Matsusaka, Mie 515-8511, Japan\bigskip }

\vspace{30pt} {\Large \scshape Abstract}
\end{center}

Quantum mechanics on sphere $S^{n}$ is studied from the viewpoint that the
Berry's connection has to appear as a topological term in the effective
action. Furthermore we show that this term is the Chern-Simons term of gauge
variables that correspond to the extra degrees of freedom of the enlarged
space.\newpage

\section{Introduction}

In recent years quantum mechanics on sphere $S^{n}$ have been studied in
various aspects\cite{McMullan:1995wz}\cite{Ohnuki:1993cb}. Interestingly,
the gauge fields are seen to emerge at the quantum level, which in turn
specify the possible inequivalent quantizations and they are coupled with a
particle constrained to move on the sphere. In ref.\cite{Ikemori:1999yf} we
have presented a picture where the wave functions are constrained to the
sphere by the ``quantum constraint''. Namely we utilize, \`{a} la Dirac, a
square root of the usual on sphere constraint. In this approach, the induced
gauge fields appear in the Hopf map from the wave function space to the real
space.

In this note we show the following :

\begin{enumerate}
\item  By considering the path integral quantization on $S^{2}$ through the
use of the wave functions mentioned above, we show that the Berry's
connection appears as a topological term in the effective action.

\item  The topological term is nothing but the Chern-Simons term of gauge
variables, which appear when we enlarge our space $S^{2}$ to $SO(3)$, by the
use of the relation $S^{2}=SO(3)/SO(2)$. The gauge variables absorb the
extra degrees of freedom of the enlarged space.

\item  We generalize the arguments in 1. and 2. to arbitrary sphere $S^{n}$.
\end{enumerate}

We recapitulate in section 2 what we have done in the previous paper\cite
{Ikemori:1999yf}. Then using the wave function introduced in section 2 we
perform the path integral quantization and obtain the effective action on $%
S^{2}$ in section 3. We shall be concerned with the cases $n=3,4$ in section
4. In section 5 we consider the general case. Section 6 is devoted to
discussions.

\section{Hopf Map and Quantization on a Sphere}

In the previous paper \cite{Ikemori:1999yf} we have quantized a system
constrained to move on a sphere by considering a square root of the ``on
sphere condition'' and have arrived at the fibre bundle structure of the
Hopf map in the cases of $S^{2}\,$and $S^{4}$. This leads to more
geometrical understanding of monopole and instanton gauge structures that
emerge in the course of quantization. We have seen that square root of the
``on sphere condition'' ${x}_{M}^{2}-r^{2}=0$ can be written as 
\begin{equation}
\left( x_{M}\Gamma _{M}-r\right) \Phi \left( \vec{x}\right) =0\ ,
\label{on sphere condition}
\end{equation}
where $\Gamma _{M}$ are the Pauli matrices for $S^{2}$ and are the Dirac
matrices for $S^{4}$ . A solution of eq.(\ref{on sphere condition}) is
expressed as 
\begin{equation}
\Phi =v\phi \ ,
\end{equation}
with $\phi $ an arbitrary complex function on the sphere. The solution to
the constraint is the space projected by $P$%
\begin{equation}
P\Phi =\Phi \ ,
\end{equation}
where the projection operator $P$ to the space spanned by $v$ is defined as 
\begin{eqnarray}
P &\equiv &vv^{\dag }\ , \\
P^{2} &=&vv^{\dag }vv^{\dag }=P\ ,  \nonumber \\
Pv &=&v\ .  \nonumber
\end{eqnarray}
The explicit form of $v$ can be written as 
\begin{equation}
v=\frac{1}{\sqrt{2r\left( r+x_{3}\right) }}\left( 
\begin{array}{c}
r+x_{3} \\ 
x_{1}+ix_{2}
\end{array}
\right) \;,  \label{v2}
\end{equation}
for $S^{2}$ and 
\begin{equation}
v=\frac{1}{\sqrt{2r\left( r+x_{5}\right) }}\left( 
\begin{array}{c}
r+x_{5} \\ 
x_{4}-i\vec{x}\cdot \vec{\sigma}
\end{array}
\right) \;,
\end{equation}
for $S^{4}$ . Then evaluation of the projected derivative $P\partial \Phi $,
through 
\begin{eqnarray}
P\partial \Phi &=&vv^{\dag }\partial \left( v\phi \right) \\
&=&vv^{\dag }\left( v\partial \phi +\partial v\phi \right)  \nonumber \\
&=&vD\phi \ ,  \nonumber
\end{eqnarray}
where 
\begin{eqnarray}
D &\equiv &\partial +A\;, \\
A &\equiv &v^{\dag }\partial v\;,  \nonumber
\end{eqnarray}
leads to the induced magnetic monopole gauge potential for $S^{2}$%
\begin{equation}
A\equiv v^{\dag }dv=\frac{i}{2r\left( r+x_{3}\right) }\left(
x_{1}dx_{2}-x_{2}dx_{1}\right) \ ,
\end{equation}
which was obtained in refs.\cite{McMullan:1995wz},\cite{Ohnuki:1993cb}, and
the instanton gauge potential for $S^{4}$%
\begin{equation}
A=i\frac{1}{2r\left( r+x_{5}\right) }\sigma _{\mu \nu }x_{\mu }dx_{\nu }\;,
\end{equation}
discussed in refs.\cite{McMullan:1994ip},\cite{Fujii:1995wn}.

\section{Path Integral Quantization on $S^{2}$}

\bigskip

According to the arguments of the previous section, the wave function on $%
S^{2}$ can be written as 
\begin{equation}
\Phi =v\phi \ .
\end{equation}
Using this wave function we calculate the transition amplitude 
\begin{eqnarray}
Z &\equiv &\langle \vec{x},v;t_{f}|\vec{x},v;t_{i}\rangle  \nonumber \\
&=&{}_{f}\langle \vec{x},v|e^{-iH(t_{f}-t_{i})}|\vec{x},v\rangle _{i} 
\nonumber \\
&=&\int [d\vec{x}]\langle v_{f}|\langle \vec{x}_{f}|e^{-iH\Delta t}|\vec{x}%
_{N-1}\rangle |v_{N-1}\rangle \langle v_{N-1}|\langle \vec{x}%
_{N-1}|e^{-iH\Delta t}|\vec{x}_{N-2}\rangle |v_{N-2}\rangle  \nonumber \\
&&\;\;\;\;\;\cdots \langle v_{1}|\langle \vec{x}_{1}|e^{-iH\Delta t}|\vec{x}%
_{i}\rangle |v_{i}\rangle \ ,
\end{eqnarray}
which by use of 
\begin{equation}
\langle v_{k}|\langle \vec{x}_{k}|e^{-iH\Delta t}|\vec{x}_{k-1}\rangle
|v_{k-1}\rangle =\int [d\vec{p}]\langle v_{k}|e^{i(p_{k}\frac{\Delta x_{k}}{%
\Delta t}-H(x_{k}))\Delta t}|v_{k-1}\rangle \;,
\end{equation}
can be written as 
\begin{equation}
\langle \vec{x},v;t_{f}|\vec{x},v;t_{i}\rangle =\int [d\vec{x}][d\vec{p}%
]e^{i\int (\vec{p}\cdot \dot{\vec{x}}-H(\vec{x}))dt}\langle
v_{f}|v_{N-1}\rangle \langle v_{N-1}|v_{N-2}\rangle \cdots \langle
v_{1}|v_{i}\rangle \;,
\end{equation}
where, for example, $[d\vec{x}]\equiv \displaystyle{\prod_{i=1}^{3}dx_{i}%
\delta (\sum_{j=1}^{3}x_{j}^{2}-}r^{2}{)}$. Furthermore, if we express 
\begin{equation}
\langle v_{k}|v_{k-1}\rangle \sim 1-\langle v|\partial _{\vec{x}}|v\rangle
\Delta \vec{x}\sim e^{i\omega }\;,  \nonumber
\end{equation}
we have 
\begin{equation}
\langle \vec{x},v;t_{f}|\vec{x},v;t_{i}\rangle =\int [d\vec{x}][d\vec{p}%
]e^{i\int (\vec{p}\cdot \dot{\vec{x}}-H(\vec{x}))dt}e^{i\oint \omega }\;,
\end{equation}
where 
\begin{equation}
\omega =i\langle v|\partial _{\vec{x}}|v\rangle \Delta \vec{x}={\frac{%
-\epsilon _{3ij}}{2r(r+x_{3})}}x_{i}\dot{x}_{j}\Delta t\equiv A_{0}\Delta
t\;.  \nonumber
\end{equation}
Consequently, to the original action 
\begin{equation}
S_{0}=\int (\vec{p}\cdot \dot{\vec{x}}-H(\vec{x}))dt\;,
\end{equation}
we have to add 
\begin{equation}
S^{\prime }={\int }A_{0}dt\;.
\end{equation}
Which implies emergence of the geometrical term added to the original
Lagrangian and the induced Lagrangian should be 
\begin{equation}
L_{S^{2}}=L_{S^{2}}^{0}+A_{0}\;.  \label{LS2}
\end{equation}

In order to see the meaning of the induced term, let us consider $H$%
-covariant formulation of Lagrangian on the coset space $G/H$ , which in our
case is $S^{2}=SO(3)/SO(2)$. We can express the original Lagrangian in terms
of $SO(3)$ variables $U=u_{0}+i\vec{\sigma}\cdot \vec{u}$ , 
\begin{equation}
L_{SO(3)/SO(2)}^{0}={r}^{2}\mathrm{Tr}[DU(DU)^{\dagger }]\;,
\label{L0SO(3)/SO(2)}
\end{equation}
where $D\equiv \displaystyle\frac{d}{dt}+i\mathcal{A}\displaystyle\frac{%
\sigma _{3}}{2}$ and $\mathcal{A}$ is a ``gauge variable'' that compensates
the redundant freedom of $SO(2)$ contained in $U$ . We can see the
equivalence of this Lagrangian with $L_{S^{2}}^{0}$ by integrating out $%
\mathcal{A}$ in eq.(\ref{L0SO(3)/SO(2)}), \textit{i.e.} 
\begin{equation}
L_{SO(3)/SO(2)}^{0}\Rightarrow {\frac{r^{2}}{4}}\mathrm{Tr}\left[ {\frac{d}{%
dt}}(U^{\dagger }\sigma _{3}U)\right] ^{2}=\frac{1}{2}\left( \dot{\vec{x}}%
\right) ^{2}=L_{S^{2}}^{0}\;,
\end{equation}
since $U^{\dagger }\sigma _{3}U=\displaystyle\frac{\vec{x}}{r}\cdot \vec{%
\sigma}$\ and $\vec{x}^{2}=r^{2}$.

The induced Lagrangian (\ref{LS2}) can be written as 
\begin{equation}
L_{SO(3)/SO(2)}=L_{SO(3)/SO(2)}^{0}+k\mathcal{A}\;,
\end{equation}
in $H$-covariant form, where $k=-1/2$ . Indeed, integrating out $\mathcal{A}$
once again we arrive at 
\begin{equation}
L_{SO(3)/SO(2)}\Rightarrow L_{S^{2}}^{0}-i\frac{1}{2}\mathrm{Tr}(\dot{U}%
U^{\dagger }\sigma _{3})-\frac{1}{8r^{2}}\;,
\end{equation}
which is equivalent to $L_{S^{2}}$ (\ref{LS2}) up to total divergence and
constant term. That is, when the system is described in terms of $SO(3)$
variables, a term proportional to the ``gauge variable'' $\mathcal{A}$,
which was introduced in order to absorb the extra degrees of freedom, is
induced. This term, which has topological origin, can be considered as a $%
(0+1)$-dimensional Chern-Simons term.

\section{$S^{3}$ and $S^{4}$}

The $S^{3}$ and $S^{4}$ cases go along the same line. In the case of $S^{4}$%
, the quantum constraint can be written as 
\begin{equation}
(x_{M}\gamma ^{M}-r)|v\rangle =0\ ,  \label{on sphere condition s4}
\end{equation}
where 
\begin{equation}
\begin{array}{ll}
\gamma ^{M} & =(\gamma ^{1},\gamma ^{2},\gamma ^{3},\gamma ^{4},\gamma
^{5})\;, \\ 
\vec{\gamma} & =\left( 
\begin{array}{ll}
0 & i\vec{\sigma} \\ 
-i\vec{\sigma} & 0
\end{array}
\right) \;,\ \gamma ^{4}=\left( 
\begin{array}{ll}
0 & 1 \\ 
1 & 0
\end{array}
\right) \;,\ \gamma ^{5}=\left( 
\begin{array}{ll}
1 & 0 \\ 
0 & -1
\end{array}
\right) \;, \\ 
\vec{\gamma} & =(\gamma ^{1},\gamma ^{2},\gamma ^{3})\;.
\end{array}
\end{equation}
Solution of the equation (\ref{on sphere condition s4}) is 
\begin{equation}
|v\rangle ={\frac{1}{\sqrt{2r(r+x_{5})}}}\left( 
\begin{array}{c}
r+x_{5} \\ 
x_{4}-i\vec{x}\cdot \vec{\sigma}
\end{array}
\right) \;,
\end{equation}
and the transition amplitude can be written as 
\begin{equation}
Z_{S^{4}}=\int [dx_{M}][dp_{M}]\exp \left[ i\int \left( p_{N}\dot{x}%
_{N}-H_{S^{4}}^{0}\right) dt\right] \mathrm{Tr}\exp \left[ \int
A_{S^{4}}^{0}dt\right] \ ,  \label{zs4}
\end{equation}
where 
\begin{equation}
A_{S^{4}}^{0}={\frac{-1}{2r(r+x_{5})}}\sigma _{\mu \nu }x_{\mu }\dot{x}_{\nu
}\ .
\end{equation}
That is, a coupling with the instanton solution is induced. As for the case
of $S^{3}$, the transition amplitude that follows from the wave function is
nothing but eq.(\ref{zs4}) with $x_{5}=0$, that is, 
\begin{equation}
Z_{S^{3}}=\int [dx_{\mu }][dp_{\mu }]\mathrm{Tr}\exp \left[ i\int \left(
p_{\nu }\dot{x}_{\nu }-H_{S^{3}}^{0}\right) dt\right] \mathrm{Tr}\exp \left[
\int A_{S^{3}}^{0}dt\right] \ ,
\end{equation}
the additional term being a coupling with the meron solution\cite
{Ikemori:1997fh} 
\begin{equation}
A_{S^{3}}^{0}={\frac{-1}{2r^{2}}}\sigma _{\mu \nu }x_{\mu }\dot{x}_{\nu }\ .
\end{equation}

As in the case for $S^{2}$ , let us consider $H$-covariant formulation of
the transition amplitude on the coset space $G/H$ , which is $SO(4)/SO(3)$
in the case for $S^3$ and is $SO(5)/SO(4)$ in the case for $S^4$. 

Let us start with the case of $S^{3}=SO(4)/SO(3)$. We write the $SO(4)$
element in the block diagonal $4\times 4$ matrix form 
\begin{equation}
SO(4)\ni G_{4}=\left( 
\begin{array}{cc}
e^{i\Theta _{\mu \nu }\sigma _{\mu \nu }} & 0 \\ 
0 & e^{i\bar{\Theta}_{\mu \nu }\bar{\sigma}_{\mu \nu }}
\end{array}
\right) \equiv \left( 
\begin{array}{cc}
g & 0 \\ 
0 & \bar{g}
\end{array}
\right) \;,
\end{equation}
where $\sigma _{ij}=\bar{\sigma}_{ji}=\frac{1}{2}\epsilon _{ijk}\sigma
_{k}\;,\;\sigma _{i4}=-\bar{\sigma}_{i4}=\frac{1}{2}\sigma _{i}$. The
Lagrangian on $S^{3}$ can be written in terms of $G_{4}$ as 
\begin{equation}
L_{SO(4)/SO(3)}^{0}={\frac{r^{2}}{2}}\mathrm{Tr}%
(G_{4}^{-1}DG_{4}(G_{4}^{-1}DG_{4})^{\dagger })\;,  \label{L0SO(4)/SO(3)}
\end{equation}
where $D\equiv \displaystyle\frac{d}{dt}+i\mathcal{A}^{SO(3)}$ and $\mathcal{%
A}^{SO(3)}$ is the gauge variable that absorbs the extra $SO(3)$ degrees of
freedom, which we write as 
\begin{equation}
\mathcal{A}^{SO(3)}\equiv \left( 
\begin{array}{cc}
\mathcal{A}_{i}\frac{\sigma _{i}}{2} & 0 \\ 
0 & \mathcal{A}_{i}\frac{\sigma _{i}}{2}
\end{array}
\right) \;.
\end{equation}
We find that the Lagrangian (\ref{L0SO(4)/SO(3)}) is equivalent to the naive
Lagrangian for a particle on $S^{3}$. Indeed, integrating out $\mathcal{A}%
^{SO(3)}$ we arrive at the following Lagrangian on $S^{3}$%
\begin{eqnarray}
L_{SO(4)/SO(3)}^{0} &\Rightarrow &-{\frac{r^{2}}{4}}\mathrm{Tr}(g\dot{g}%
^{-1}-\bar{g}\dot{\bar{g}}^{-1})^{2}  \nonumber \\
&=&{\frac{r^{2}}{4}}\mathrm{Tr}\left[ {\frac{d}{dt}}{(\bar{g}^{-1}g)}{\frac{d%
}{dt}}{(g^{-1}\bar{g})}\right] \\
&=&{\frac{r^{2}}{4}}\mathrm{Tr}\dot{Q}_{3}\dot{Q}_{3}^{-1}=\frac{1}{2}\dot{x}%
_{\mu }^{2}=L_{S^{3}}^{0}\;,  \nonumber
\end{eqnarray}
where $Q_{3}\equiv \bar{g}^{-1}g=\displaystyle\frac{x_{4}}{r}+i\displaystyle%
\frac{x_{i}}{r}\sigma _{i}$ and $\displaystyle\sum_{\mu =1}^{4}x_{\mu
}^{2}=r^{2}\;.$

Next we add a term $\mathrm{Tr}(K\mathcal{A}^{SO(3)})$ to the Lagrangian,
where the constant $K$ is given by 
\begin{equation}
K=\left( 
\begin{array}{cc}
k_{i}\sigma _{i} & 0 \\ 
0 & k_{i}\sigma _{i}
\end{array}
\right) \;.
\end{equation}
Then, integrating out $\mathcal{A}^{SO(3)}$, we find that the Lagrangian, 
\begin{equation}
L_{SO(4)/SO(3)}=L_{SO(4)/SO(3)}^{0}+\mathrm{Tr}(K\mathcal{A}^{SO(3)})\ ,
\end{equation}
goes to 
\begin{equation}
L_{SO(4)/SO(3)}\Rightarrow L_{S^{3}}^{0}-i\mathrm{Tr}\left[ G_{4}\dot{G}%
_{4}^{-1}K\right] -{\frac{1}{2r^{2}}}\mathrm{Tr}(K^{2})\ .
\end{equation}
Furthermore, we can show that the transition amplitude $Z_{SO(4)/SO(3)}$
corresponding to this Lagrangian is identical to $Z_{S^{3}}$. As we can write%
%
%
%
%
%
%
%
%
%
%
\begin{equation}
G_{4}=h^{-1}\left( 
\begin{array}{cc}
\bar{g}^{-1}g & 0 \\ 
0 & 1
\end{array}
\right) \;,
\end{equation}
we have 
\begin{equation}
G_{4}\dot{G}_{4}^{-1}=h^{-1}(-4iA_{S^{3}}^{0})h+h^{-1}\dot{h}\;,
\end{equation}
where $h$ is an $SO(3)\;$element written in $4\times 4$ block diagonal
matrix form. Thus the Lagrangian is expressed as 
\begin{equation}
L_{SO(4)/SO(3)}\Rightarrow L_{S^{3}}^{0}-i\mathrm{Tr}(Kh^{-1}\dot{h})-4%
\mathrm{Tr}(hKh^{-1}A_{S^{3}}^{0})\;+\mathrm{const.\;}.
\end{equation}
We note that this Lagrangian has the same form as that obtained in ref.\cite
{McMullan:1995wz}. Here if we define 
\begin{equation}
\left( 
\begin{array}{cc}
S^{i}\sigma _{i} & 0 \\ 
0 & S^{i}\sigma _{i}
\end{array}
\right) \equiv \frac{1}{2}h^{-1}\left( 
\begin{array}{cc}
\sigma _{3} & 0 \\ 
0 & \sigma _{3}
\end{array}
\right) h\;,
\end{equation}
the corresponding Hamiltonian can be written as%
\begin{equation}
H=H_{S^{3}}^{0}-2S^{i}(A_{S^{3}}^{0})^{i}\;,
\end{equation}
where $S^{i}$ is a spin variable that satisfies $[S^{i},S^{j}]=i\epsilon
^{ijk}S^{k}\;.\;$Based on this Hamiltonian with $k_{i}=-\frac{1}{4}\delta
_{i3},$ we derive the transition amplitude by integrating the spin degrees
of freedom\cite{Nielsen:1988sa}\cite{Kashiwa:1990pk}, 
\begin{equation}
Z_{SO(4)/SO(3)}=\int [dx_{\mu }][dp_{\mu }]\exp \left[ i\int \left( p_{\nu }%
\dot{x}_{\nu }-H_{S^{3}}^{0}\right) dt\right] \mathrm{Tr}\exp \left[ i\int
A_{S^{3}}^{0}dt\right] \;.
\end{equation}
We have confirmed this equation under the gauge condition $%
(A_{S^{3}}^{0})^{1}=(A_{S^{3}}^{0})^{2}=0\;.$ This expression coincides
completely with the previous $Z_{S^{3}}\;.$%


Next we turn to the discussion on $S^{4}$. The naive Lagrangian for the
particle on $S^{4}$ in terms of $SO(5)$ variable $G_{5}$ can be written as 
\begin{equation}
L_{SO(5)/SO(4)}^{0}=\frac{{r}^{2}}{2}\mathrm{Tr}\left(
G_{5}^{-1}DG_{5}\left( G_{5}^{-1}DG_{5}\right) ^{\dag }\right) \;,
\end{equation}
where $D\equiv \displaystyle\frac{d}{dt}+i\mathcal{A}^{SO(4)}$ is the $SO(4)$%
-covariant derivative. By integrating out $\mathcal{A}^{SO(4)}$, this
Lagrangian can be seen to be equivalent to $L_{S^{4}}^{0}$. This may be
explicitly shown in the representation such as 
\begin{equation}
\mathcal{A}^{SO(4)}=\left( 
\begin{array}{cc}
\mathcal{A}_{\mu \nu }\sigma _{\mu \nu } & 0 \\ 
0 & \mathcal{A}_{\mu \nu }\bar{\sigma}_{\mu \nu }
\end{array}
\right) =\left( 
\begin{array}{cc}
\mathcal{A}_{+} & 0 \\ 
0 & \mathcal{A}_{-}
\end{array}
\right) \equiv \mathcal{A}_{a}^{SO(4)}T_{a}\ ,
\end{equation}
where $T_{a}$ is the generator corresponding to the $SO(4)$ subgroup of $%
SO(5)$. We separate as $G_{5}\dot{G}_{5}^{-1}\equiv ig_{\bot
}^{a}T_{a}+ig_{\Vert }^{\alpha }T_{\alpha }\equiv G_{\bot }+G_{\Vert }$ ( $%
T_{\alpha }$ is the remaining generator of $SO(5)$ ). Integrating out $%
\mathcal{A}^{SO(4)}$, we see 
\begin{equation}
L_{SO(5)/SO(4)}^{0}\Rightarrow -{\frac{r^{2}}{2}}\mathrm{Tr}G_{\Vert }^{2}=%
\frac{1}{2}\dot{x}_{M}^{2}=L_{S^{4}}^{0}\;,
\end{equation}
where $G_{5}^{-1}\gamma ^{5}G_{5}=\displaystyle\sum_{M=1}^{5}\displaystyle%
\frac{x_{M}}{r}\gamma _{M}\;$and$\;\displaystyle%
\sum_{M=1}^{5}x_{M}^{2}=r^{2} $.

Next we assume that the term $\mathrm{Tr}(K\mathcal{A}^{SO(4)})$ has been
induced to the system on $S^{4}$ , where the constant $K$ is the algebra of $%
SO(4)$, the general form of which is given by 
\begin{equation}
K=\left( 
\begin{array}{cc}
K_{+} & 0 \\ 
0 & K_{-}
\end{array}
\right) =K_{a}T_{a}\;.
\end{equation}
Then integrating out $\mathcal{A}^{SO(4)}$ we obtain the Lagrangian on $S^{4}
$ 
\begin{equation}
L_{SO(5)/SO(4)}\Rightarrow L_{S^{4}}^{0}-i\mathrm{Tr}(G_{5}\dot{G}%
_{5}^{-1}K)-\frac{1}{2r^{2}}\mathrm{Tr}(K^{2})\;.  \label{LSO(5)/SO(4)}
\end{equation}
As in the case of $S^{3}$, we calculate transition amplitude by means of the
Hamiltonian 
\begin{equation}
H=H_{S^{4}}^{0}+2S^{i}(A_{S^{4}}^{0})^{i}\;,
\end{equation}
which corresponds to the effective Lagrangian (\ref{LSO(5)/SO(4)}), and
integrate out the spin variables $S^{i}$ defined by 
\begin{equation}
\left( 
\begin{array}{cc}
0 & 0 \\ 
0 & S^{i}\sigma _{i}
\end{array}
\right) \equiv \frac{1}{2}\tilde{h}^{-1}\left( 
\begin{array}{cc}
0 & 0 \\ 
0 & \sigma _{3}
\end{array}
\right) \tilde{h}\;,
\end{equation}
where $\tilde{h}$ is an element of the $SO(4)$ subgroup. Then, we arrive at 
\begin{equation}
{\ }Z_{SO(5)/SO(4)}=\int [dx_{M}][dp_{M}]\exp \left[ i\int \left( p_{N}\dot{x%
}_{N}-H_{S^{4}}^{0}\right) dt\right] \mathrm{Tr}\exp \left[ i\int
A_{S^{4}}^{0}dt\right] \;,
\end{equation}
where the choice has been used of $K_{+}=0\;,\;K_{-}=-\frac{1}{2}\sigma
_{3}\;.$

Thus, we see that for both cases $S^{3}\;$and $S^{4}$, the $(0+1)$%
-dimensional Chern-Simons terms, that are proportional to the gauge
variable, are induced in the course of quantization.%
\quad%
%

\section{$S^{n}$}

\bigskip What has been argued so far can be generalized to all $n$ . First,
we obtain the transition amplitude for a particle on $S^{n}$ that follows
from the wave function with the ``on sphere condition'' taken into account.
In order to do this, we consider the cases $n=2m$ and $n=2m-1$ separately.
We start with $n=2m$ case and prepare $2^{m}\times 2^{m}$ matrices $\Gamma
_{N}^{(2m)}\;\left( N=1,2,\cdots ,2m+1\right) $

\begin{equation}
\{\Gamma _{M}^{(2m)},\Gamma _{N}^{(2m)}\}=2\delta _{MN},\qquad \lbrack
\Gamma _{M}^{(2m)},\Gamma _{N}^{(2m)}]=2i\Sigma _{MN}^{(2m)}\;.
\end{equation}
The explicit form of $\Gamma _{N}^{(2m)}$ can be obtained as 
\begin{eqnarray}
\Gamma _{i}^{(2m)} &=&\sigma _{2}\otimes \Gamma _{i}^{(2m-2)}\qquad
(i=1,\cdots ,2m-1)\;,  \nonumber \\
\Gamma _{2m}^{(2m)} &=&\sigma _{1}\otimes 1\;, \\
\Gamma _{2m+1}^{(2m)} &=&\sigma _{3}\otimes 1\;,  \nonumber
\end{eqnarray}
and $SO(2m)$ generators $\Sigma _{\mu \nu }^{(2m)}\quad (\mu ,\nu
=1,2,\cdots ,2m)$ are expressed in block diagonal form as 
\begin{equation}
\left( 
\begin{array}{cc}
\Sigma _{\mu \nu }^{(2m)+} & 0 \\ 
0 & \Sigma _{\mu \nu }^{(2m)-}
\end{array}
\right) \;.
\end{equation}
The spinor wave function $|v\rangle $ in $2^{m}\times 2^{m-1}$ matrix form
is constrained to satisfy the ``on sphere condition'' 
\begin{equation}
(x_{N}\Gamma _{N}^{(2m)}-r)|v\rangle =0\;.
\end{equation}
The non-trivial solution to the equation is given as 
\begin{equation}
|v\rangle ={\frac{1}{\sqrt{2r(r+x_{2m+1})}}}\left( 
\begin{array}{c}
r+x_{2m+1} \\ 
x_{2m}-ix_{i}\Gamma _{i}^{(2m-2)}
\end{array}
\right) \;.
\end{equation}
With the help of $|v\rangle $ we perform the path integral to obtain the
transition amplitude, 
\begin{equation}
Z_{S^{2m}}=\int [dx_{M}][dp_{M}]\exp \left[ i\int (p_{N}\dot{x}%
_{N}-H_{S^{2m}}^{0})dt\right] \mathrm{Tr}\exp \left[ i\int A_{S^{2m}}^{0}dt%
\right] \;,  \label{zs2m}
\end{equation}
where 
\begin{equation}
A_{S^{2m}}^{0}={\frac{-1}{2r(r+x_{2m+1})}}\Sigma _{\mu \nu }^{(2m)+}x_{\mu }%
\dot{x}_{\nu }\;.
\end{equation}
Thus, \ we can claim that a coupling to the generalized instanton
configuration is induced \cite{Fujii:1995wn}.

The transition amplitude for a particle on $S^{2m-1}$ is nothing but eq.(\ref
{zs2m}) with $x_{2m+1}=0$. That is, 
\begin{equation}
Z_{S^{2m-1}}=\int [dx_{\mu }][dp_{\mu }]\exp \left[ i\int (p_{\nu }\dot{x}%
_{\nu }-H_{S^{2m-1}}^{0})dt\right] \mathrm{Tr}\exp \left[ i\int
A_{S^{2m-1}}^{0}dt\right] \;,
\end{equation}
The additional term is a coupling to the generalized meron solution in
arbitrary odd dimensions\cite{Ikemori:1997fh}, 
\begin{equation}
A_{S^{2m-1}}^{0}={\frac{-1}{2r^{2}}}\Sigma _{\mu \nu }^{(2m)+}x_{\mu }\dot{x}%
_{\nu }\;.
\end{equation}

Next, exactly as in the previous discussions, having in mind that $%
S^{n}=SO(n+1)/SO(n)$ we can describe the system using the elements of $%
SO(n+1)$. This suggest that the induced term, which was obtained through the
above mentioned path integration, appears as a term proportional to the
``gauge variable'' $\mathcal{A}^{SO(n)}$ that was introduced in order to
absorb the extra degrees of freedom. Thus, for the description of quantum
mechanics on $S^{n}$ using the $SO(n+1)$ variables, the gauge variable $%
\mathcal{A}^{SO(n)}$ is expected to be induced. Namely, we claim that the
induced term is a term proportional to the ``gauge variable'' also for the
general dimension $n$.

\section{Discussions}

We have seen that, when describing quantum mechanics on the sphere in terms
of the wave functions that satisfy the ``square root'' of\ ``on sphere
condition'', a particular gauge configuration appears in the transition
amplitude due to the geometrical reasons. These results are consistent with
those obtained in ref.\cite{Tanimura:1996cg} from a different point of view.
These configurations are monopoles and (generalized) instantons for even
dimensional spheres and (generalized) merons for odd dimensional spheres.

Furthermore, we have shown in this note that it is possible to interpret
this situation as an induction of a term proportional to the ``gauge
variable'' that was introduced in order to absorb the extra degrees of
freedom, when we describe the $S^{n}$ system in terms of the $SO(n+1)$
variables according to the relation $S^{n}=SO(n+1)/SO(n)$. The induced terms
have topological meaning and can be considered as a Chern-Simons terms in $%
0+1$ dimensions.

If we extend this result, obtained in quantum mechanics, to the field
theories where the fields are constrained to the sphere, we could expect an
induction of Chern-Simons gauge fields.\ For example, for $O(3)$ sigma model
in $2+1$ dimensions, fields are on $S^{2}=SU(2)/U(1)$ and we expect $U(1)$
Chern-Simons term to be induced in this case. This possibility was also
suggested in refs.\cite{Igarashi:1997tk},\cite{Kobayashi:1997rf}.


\end{document}